\documentclass[12pt]{iopart}
\usepackage{graphicx}
\usepackage{dcolumn}
\usepackage{bm}

\begin{document}

\title[Order in de Broglie - Bohm quantum mechanics]
{Order in de Broglie - Bohm quantum mechanics}

\author{G. Contopoulos, N.Delis and C. Efthymiopoulos}

\address{Research Center for Astronomy and Applied Mathematics,
Academy of Athens, Soranou Efesiou 4, 11527 Athens, Greece}
\ead{gcontop@academyofathens.gr, nikdelis@sch.gr,
cefthim@academyofathens.gr}

\begin{abstract}
A usual assumption in the so-called
{\it de Broglie - Bohm} approach to quantum dynamics is that the
quantum trajectories subject to typical `guiding' wavefunctions
turn to be quite irregular, i.e. {\it chaotic} (in the
dynamical systems' sense). In the present paper, we consider
mainly cases in which the quantum trajectories are {\it ordered},
i.e. they have zero Lyapunov characteristic numbers. We use
perturbative methods to establish the existence of such trajectories
from a theoretical point of view, while we analyze their properties
via numerical experiments. Using a 2D harmonic oscillator system,
we first establish conditions under which a trajectory can be shown
to avoid close encounters with a moving nodal point, thus avoiding
the source of chaos in this system. We then consider series expansions
for trajectories both in the interior and the exterior of the domain
covered by nodal lines, probing the domain of convergence as well as how
successful the series are in comparison with numerical computations
or regular trajectories. We then examine a H\'{e}non - Heiles system
possessing regular trajectories, thus generalizing previous results.
Finally, we explore a key issue of physical interest in the context
of the de Broglie - Bohm formalism, namely the influence of order
in the so-called {\it quantum relaxation} effect. We show that
the existence of regular trajectories poses restrictions to the quantum
relaxation process, and we give examples in which the relaxation
is suppressed even when we consider initial ensembles of only
chaotic trajectories, provided, however, that the system as a whole
is characterized by a certain degree of order.
\end{abstract}


\maketitle

\section{Introduction}

The de Broglie - Bohm version of quantum mechanics \cite{debro1928}\cite{bohm1952}\cite{bohmhil1993}\cite{hol1993}\cite{durteu2009}
considers trajectories guided by the wave function, which is a
solution of Shr\"{o}dinger's equation, i.e. (in the one particle case)
\begin{equation}
\bigg(-\frac{1}{2}\nabla^2+V\bigg)\psi=i\frac{\partial\psi}{\partial
t}
\end{equation}
where $V$ is the potential and we have set $m=\hbar=1$. The de Broglie
- Bohm mechanics gives results which are equivalent to the Copenhagen
version of quantum mechanics. However it gives further information
about the trajectories of particles, or fluid elements of a probabilistic
fluid, given by the probability $p=|\psi|^2$ \cite{mad1926}.
The de Broglie - Bohm equations of motion are
\begin{equation}
\frac{d\vec{r}}{dt}=Im(\frac{\nabla\psi}{\psi})~~.
\end{equation}

The quantum trajectories are either ordered or chaotic. The problem
of chaos in the de Broglie - Bohm theory has been studied extensively
(indicative references are
\cite{duretal1992}\cite{faisch1995}\cite{parval1995}\cite{depol1996}
\cite{dewmal1996}\cite{iacpet1996}\cite{fri1997}\cite{konmak1998}
\cite{wuspru1999}\cite{maketal2000}\cite{cush2000}\cite{desalflo2003}
\cite{falfon2003}\cite{wispuj2005}\cite{wisetal2007}\cite{schfor2008}).
In four previous papers \cite{eftcon2006}\cite{eftetal2007}\cite{coneft2008})
\cite{eftetal2009} we studied in detail the basic mechanism by
which chaos is generated when the trajectories come close to {\it nodal
points}, where $\psi=0$. The appearance of {\it quantum vortices} at
nodal points is a well known phenomenon \cite{dir1931}, which, besides
chaos, leads to a number of interesting consequences for quantum
dynamics in general (e.g. \cite{mccwya1971}\cite{hiretal1974a}
\cite{hiretal1974b}\cite{sanzetal2004a}\cite{sanzetal2004b}).
Quantitative studies of chaos related to the
effects of vortices are presented in \cite{fri1997}\cite{schfor2008}
\cite{eftetal2009}. The fact that the generation of chaos is due
to the {\it motion} of quantum vortices was first pointed out in
\cite{wispuj2005}(see also \cite{wisetal2007}). Our own main result
was to show that chaos is due to the formation (by moving nodal points)
of `nodal point - X-point complexes'
\cite{eftetal2007}\cite{coneft2008}\cite{eftetal2009}.
More specifically, we found that near every nodal
point there is a hyperbolic point, called X-point, that has two
unstable directions, opposite to each other, and two stable
directions, again opposite to each other. Chaos is introduced when
a trajectory approaches such an X-point. Furthermore chaos is stronger
when the X-point is closer to the nodal point. As shown in \cite{eftetal2009},
the existence of an X-point close to a nodal point is a consequence
of the topological properties of the quantum flow holding in the
neighborhood of nodal points for an arbitrary form of the wavefunction.
Furthermore, we made a theoretical analysis of the dependence of the
Lyapunov characteristic numbers of the quantum trajectories on the
size and speed of the quantum vortices. Roughly, the local value of
Lyapunov characteristic numbers scales inversely proportionally to
the distance of a quantum trajectory from the stable manifold of the X-point,
and to the speed of the nodal point. Furthermore, the size of the nodal
point - X-point complex scales inversely proportionally to the speed of
the nodal point. These properties were used to estimate how close a
trajectory must approach the X-point for chaos to become effective,
thus explaining numerical results found in \cite{eftetal2007} and
\cite{coneft2008}.

On the other hand trajectories that never approach a "nodal point -
X point complex" are ordered. Such trajectories were considered in
\cite{eftetal2007} (called hereafter EKC). In that paper
we found the trajectories and the nodal lines (i.e. the lines formed
by moving nodal points) in the case of a simple model of a 2D harmonic
oscillator, when one considers a superposition of three eigenfunctions.
Other examples of ordered orbits, following a similar analysis, were
given in \cite{eftetal2009}.

In EKC we found the forms of the nodal lines and some limiting
curves that are never crossed by the nodal lines. In fact we found
that the nodal lines never come (a) close to the center or close to
the axis $x=0$ (b) close to the axis $y=0$ for large absolute values
of $x$, and (c) large values of both $x$ and $y$. This is exemplified
in Fig.\ref{haordcha}a,b, referring to an example of trajectory
calculations taken from EKC (see section 2 below for details).
The main effect to be noted is that there are empty regions left
by the nodal lines. As shown in EKC and in \cite{eftetal2009},
the appearance of order is due to the presence of such regions
devoid of nodal points at all times $t$. Further examples in a
different (square-box) model were given in \cite{conetal2008}.

The most notable physical consequence of the existence of regular
trajectories is related to a phenomenon called {\it quantum relaxation}
\cite{val1991}\cite{valwes2005}. If we allow the initial conditions
of the Bohmian particles to obey a probability distribution
$p_0\neq|\psi_0|^2$, where $\psi_0$ is the initial wavefunction, then,
under some conditions we find that $p_t$ approaches $|\psi_t|^2$
asymptotically, at a coarse-grained level, as the time  $t$
increases. The possibility to obtain an asymptotic approach of
$p$ to $|\psi|^2$ due to stochastic fluctuations of the motion
of the probabilistic fluid elements was first pointed out by
Bohm and Vigier \cite{bohmvig1954}. However, the theory of
Valentini \cite{val1991} does not rely on extra assumptions
besides the absence of the so-called micro-state fine structure
in the initial distribution of the particles. Quantum relaxation
has been observed in a number of numerical simulations
(\cite{valwes2005}\cite{eftcon2006}\cite{towetal2011}).
The main condition for its effectiveness, however, is that
the trajectories considered must be {\it chaotic}. In fact,
it is known that the existence of ordered trajectories limits
the effectiveness of quantum relaxation \cite{eftcon2006}.

Since the physics of quantum relaxation has attracted recent
interest in a variety of physical contexts
(e.g. \cite{ben2010}\cite{colstru2010}\cite{val2010}),
it is important to understand, besides the mechanisms of chaos,
also the mechanisms by which order appears in de Broglie - Bohm mechanics.
This is our purpose in the present paper.

The paper is organized as follows: In section 2,
we consider inner ordered trajectories, i.e. ordered
trajectories appearing in the central region of the configuration space
of the same model as in EKC. In particular, using perturbation series
we explain why there can be some ordered trajectories partly overlapping
with the domain covered by nodal lines in this model. In section
3 we discuss the issue of convergence of the perturbation series
representing the regular trajectories. Section 4 deals with ordered
trajectories far from the center, i.e. beyond the domain covered by
nodal lines. In section 5 we discuss ordered trajectories in a more
general (H\'{e}non - Heiles) model than the 2D harmonic oscillator.
Section 6 deals with the impact of order on the effect of quantum
relaxation, while section 7 discusses further examples and some
quantitative analysis of the same subject. Finally, section 8
presents the main conclusions of the present study.

\section{Inner ordered trajectories}
In EKC we considered the Bohmian trajectories in the 2D harmonic
oscillator model:
\begin{equation}\label{ham2dharm}
H={1\over 2}(p_x^2+p_y^2) + {1\over 2}(x^2 + (c y)^2)
\end{equation}
when the guiding field is the superposition of the ground state
and the two first excited states
\begin{equation}\label{eigenharm}
\psi(x,y,t) = e^{-{x^2+cy^2\over 2}-i{(1+c)t\over 2}}\big(
1+axe^{-it}+bc^{1/2}xye^{-i(1+c)t}\big)~~
\end{equation}
with real amplitudes $a,b$ and incommensurable frequencies
$\omega_1=1$, $\omega_2=c$. The equations of motion are:
\begin{eqnarray}\label{eqmo}
{dx\over dt}&=&-{a \sin t + bc^{1/2}y \sin(1+c)t \over G}\\
{dy\over dt}&=&-{bc^{1/2}x
\left(ax\sin ct +\sin(1+c)t\right) \over G}\nonumber
\end{eqnarray}
where
\begin{equation}\label{gi}
G=1+a^2x^2+b^2cx^2y^2+2ax\cos t+ 2bc^{1/2}xy\cos(1+c)t
+ 2abc^{1/2}x^2y\cos ct
\end{equation}

As shown in EKC, the domains covered by the nodal lines are
hyperbola-like, leaving both inner and outer domains devoid of
nodal points. The nodal points are at
\begin{equation}\label{nodal}
x_N=-{\sin(1+c)t\over a\sin ct},
~~~y_N=-{a\sin t\over bc^{1/2}\sin(1+c)t}
\end{equation}

\begin{figure}
\centering
\includegraphics[scale=0.6]{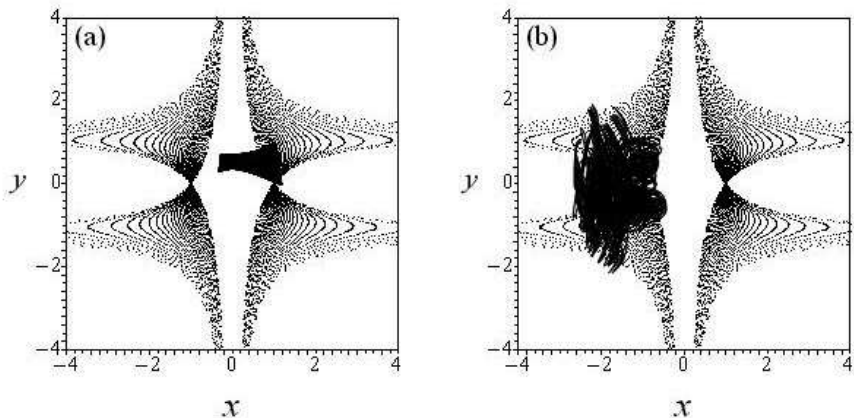}
\caption{Bohmian trajectories in a 2D harmonic oscillator model when the
wavefunction $\psi$ is a superposition of three states (Eq.(\ref{eigenharm})),
and $a=b=1$, $c=\sqrt{2}/2$. (a) An ordered trajectory with $x_0=y_0=1$,
and (b) a chaotic trajectory with $x_0=y_0=-1$. In the background we
mark the nodal lines, i.e. the solutions for $x,y$ of the equation
$\psi(x,y,t)=0$. }
\label{haordcha}
\end{figure}
Trajectories that never approach the `nodal point - X point complex'
are ordered. Such trajectories were found in EKC both numerically and
by analytical expansions. We can explicitly show that whether an orbit
is regular or chaotic depends on how close the orbit can come to the
nodal point. Furthermore, we find that almost symmetric initial
conditions can give rise to quite different trajectories (ordered
or chaotic), because the overlapping of the initial conditions
with the domain of the nodal lines does not guarantee that there
is a close encounter (in both space a time) of the trajectories
with the nodal point.

As an example, trajectories starting on the upper right quadrant are
regular, even if their initial conditions are located in a region
occupied by the nodal lines (Fig.\ref{haordcha}a, case $x_0=y_0=1$),
while trajectories starting in the lower left region occupied by the nodal
lines are chaotic (Fig.\ref{haordcha}b, case $x_0=y_0=-1$). The reason for
this difference is that, when the moving point is close to its maximum ($x$
and $y$), it can be shown that the nodal point is always far from the
moving point in the first case ($x_0=y_0=1$), while it comes several
times close to it in the second case ($x_0=y_0=-1$). A rough estimate
of the position of the moving point can be given by a first order
theory. Namely, from Eqs.(\ref{eqmo}) we derive up to first order in $a,b$:
\begin{equation}\label{solx1}
x=x_0+a(\cos t-1)+{bc^{1/2}y_0\over 1+c}\bigg(\cos(1+c)t-1\bigg)+\ldots
\end{equation}
\begin{equation}\label{soly1}
y=y_0+{bc^{1/2}x_0\over 1+c}\bigg(\cos(1+c)t-1\bigg)+\ldots
\end{equation}
These expressions are found in practice to be fairly accurate
well beyond the domain (in $a,b$) where the convergence of the series
representing $x$ or $y$ can be rigorously established (see section 3
below), and for initial conditions $x_0,y_0$ satisfying
$|x_0y_0|<1/(bc^{1/2})$ (see EKC). Using the approximate expression
(\ref{solx1}), we find that we have a maximum $x=x_0$ when
\begin{equation}\label{max1}
a\sin t+bc^{1/2}y_0\sin(1+c)t~=~0,
\end{equation}
provided that the time $t$ specified by (\ref{max1}) satisfies
the inequality
\begin{equation}\label{max2}
a\cos t+bc^{1/2}y_0(1+c)\cos(1+c)t~>~0~~.
\end{equation}
Equation (\ref{max1}) gives the times of the maximum $x$.
If the quantity
\begin{equation}\label{q}
q={bc^{1/2}\over a}y_0
\end{equation}
is relatively small (e.g. $|q|\leq 0.5$) the times of maximum $t$
are close to $2K\pi$ for some integer $K$. Then, we can write
\begin{equation}\label{tmax}
t=2K\pi+t'
\end{equation}
with
\begin{equation}\label{sintmax}
\sin t = \sin t'\approx -q\sin[(1+c)2K\pi] = -q\sin (c2K\pi)
\end{equation}
and the inequality (\ref{max2}) is satisfied.

At such a time the nodal point is near the position
\begin{equation}\label{xnmax}
x_N=-{\sin(1+c)t\over a\sin ct}={\sin t\over qa\sin ct}\approx-{1\over a}
\end{equation}
\begin{equation}\label{ynmax}
y_N=-{a\sin t\over bc^{1/2}\sin(1+c)t}\approx y_0
\end{equation}

\begin{figure}
\centering
\includegraphics[scale=0.6]{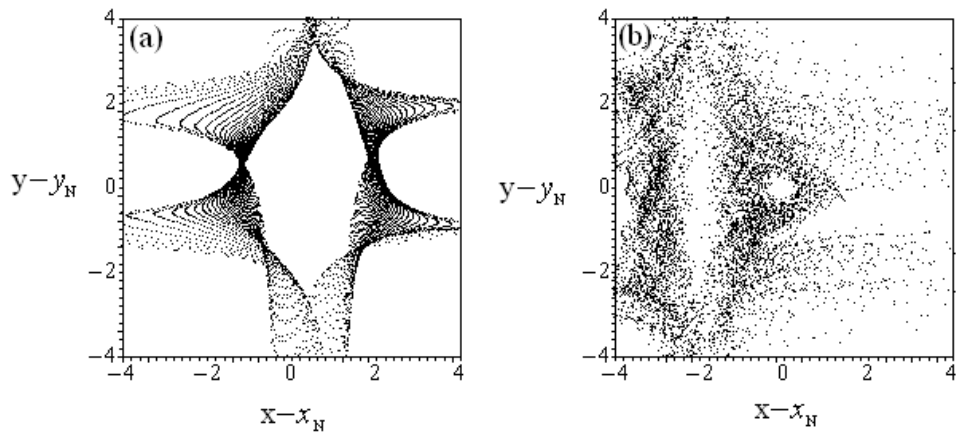}
\caption{Distribution of the quantities $x(t)-x_N(t)$, $y(t)-y_N(t)$,
where $x(t),y(t)$ are a trajectory's coordinates and $x_N(t),y_N(t)$
are the nodal point coordinates at various times $t$, (a) for the
ordered trajectory, and (b) for the chaotic trajectory of
Fig.\ref{haordcha}. }
\label{haordchand}
\end{figure}
Therefore when $x$ is at its maximum the nodal point is at a negative $x_N$.
Thus if $x_0>0$, $y_0>0$ the moving point is in the upper right quadrant,
while the nodal point ($x_N,y_N$) is in the upper left quadrant, i.e. far
from the moving point. On the other hand if $x_0<0$, $y_0<0$ the moving point
at its maximum $x$ is in the lower left quadrant, and the nodal point is also
in the lower left quadrant, therefore the two points can be close and this
closeness introduces chaos.

The difference between the ordered case $x_0=y_0=1$ and the chaotic case
$x_0=y_0=-1$ can be seen in Figs.\ref{haordchand}a,b where we give the
points ($u=x_0-x_N$, $v=y_0-y_N$) whenever the distance $d_N=\sqrt{u^2+v^2}$
goes through a local minimum in time. In this figure the nodal point is at
$u_N=v_N=0$. We see that in the first case the local minima $d_N$ in time
never approach the value $0$, while in the second case $d_N$ sometimes
approaches zero. This explains why the first trajectory is ordered,
while the second trajectory is chaotic.

\section{Series convergence}
The series representing regular orbits can be convergent only within a
certain domain in the space of parameters ($a,b$ in the case of the model
(\ref{eigenharm})), and/or in the initial conditions $x_0,y_0$. However,
in some simple cases we can show that there are regular orbits even beyond
the domains of convergence of the series constructed to represent such
orbits.

One such example is found when $b=0$. In this case all the trajectories
in the model (\ref{eigenharm}) are periodic, with period $T=2\pi$, thus
they are regular. Such trajectories were considered in \cite{coneft2008}.
The equations of motion take a simple form:
\begin{equation}\label{eqper}
{dx\over dt}=-{a\sin t\over 1+a^2x^2+2a x \cos t},~~~{dy\over dt}=0~~~.
\end{equation}
Thus, all the trajectories are straight horizontal lines, i.e.
$y(t)=y_0=\mbox{const.}$.

\begin{figure}
\centering
\includegraphics[scale=0.45]{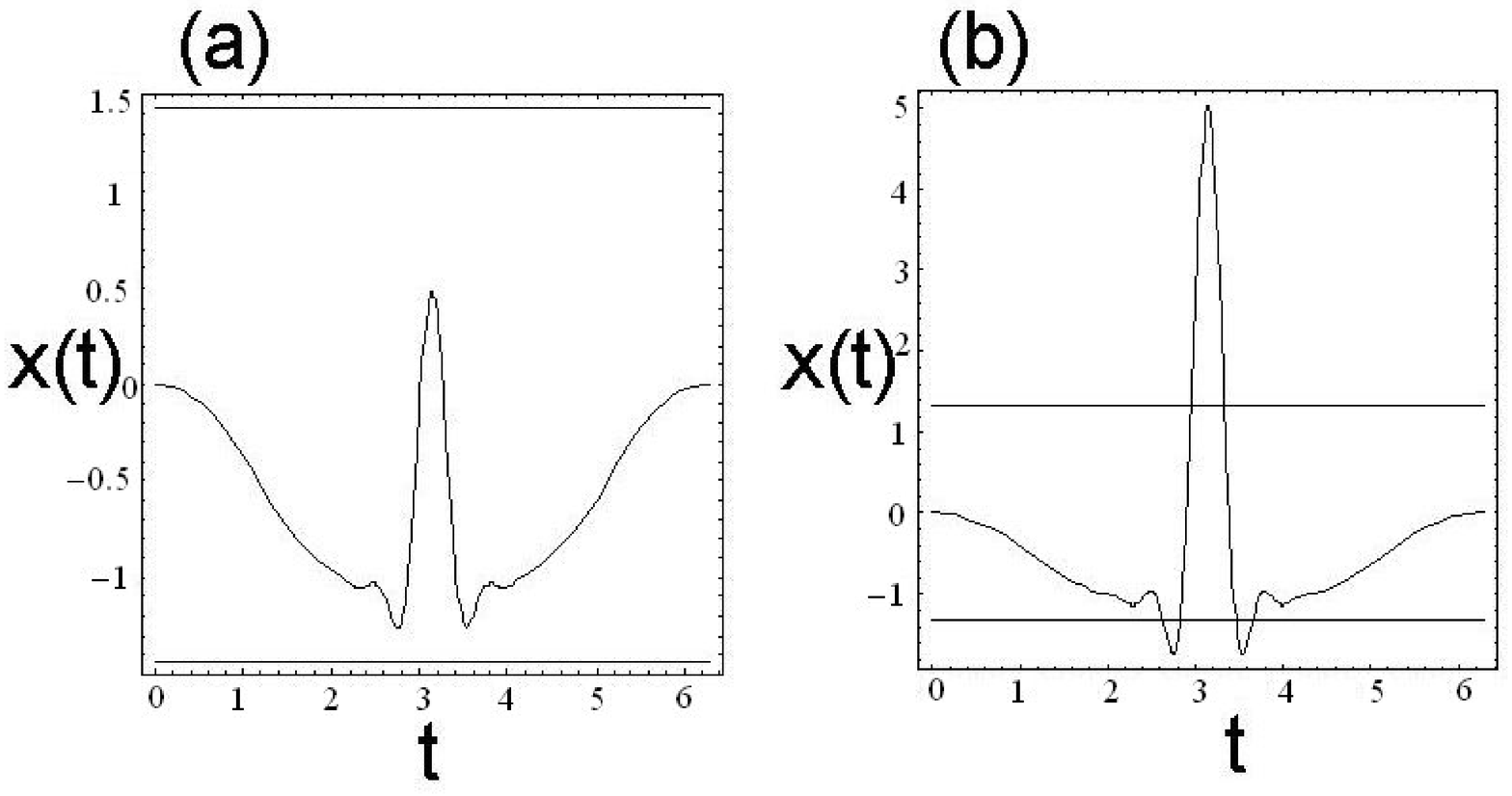}
\caption{The time evolution of x(t) in a series representation
(truncated at order 20) of the trajectory with initial conditions
$x_0=y_0=0$ when $c=\sqrt{2}/2$, $b=0$ and (a) $a=0.7$,
(b) $a=0.75$. The two horizontal lines in each plot mark
the interval of convergence of the series in the x-axis.
The crossing of these lines by the series solution for
$x(t)$ indicates that the series are divergent.}
\label{haserconv}
\end{figure}
Now, a series representation for the solution $x(t)$ can be obtained as
a limiting case of the expansion considered in section 2. We
set $x(t)=x_0 + x_1(t)+x_2(t)+...$, where $x_k(t)$ is of order $k$
in the parameter $a$. For any fixed value of $x,t$, such a series
is valid in the domain of analyticity of the function appearing
in the r.h.s. of Eq.(\ref{eqper}) with respect to $a$ (considered
as a complex variable). The latter domain can be determined by
finding the nearest pole (with respect to the origin) of the
function $1/G$, where $G=1+a^2x^2+2a x \cos t$. For fixed $(x,t)$,
the solution of $G=0$ is
\begin{equation}
a={-\cos t \pm i|\sin t|\over x}
\end{equation}
Thus the nearest pole is at the distance $|a|=1/|x|$ for all
times $t$.

The latter property allows for an easy numerical convergence test
of the series representing $x(t)$. I.e., for a fixed value of $a$,
we compute a finite truncation
\begin{equation}
x(t)\simeq x_0+\sum_{k=1}^N x_k(t)
\end{equation}
where $N$ is a (finite) truncation order, and check whether the
numerically computed values of $x(t)$ by the above expression
ever cross the limit $|x(t)a|<1$ in the interval $0\leq t<2\pi$.
Figure 3 shows this criterion, when $x_0=0$, $N=20$, and $a$
is given two different values, namely $a=0.7$ (Fig.3a) and
$a=0.75$ (Fig.3b). The horizontal straight
lines correspond to the limit $|x(t)a|=1$. The crossing of this
limit takes place a little beyond $a=0.7$, thus we conclude that
the series representing periodic trajectories are divergent beyond this
value of $a$. It should be stressed, however, that the fact that
the series are divergent only implies that the method used to
compute these series no longer yields a useful representation
of the true trajectories, since, in fact, the true trajectories
{\it are} regular (because they are periodic).

The same effect is found when $b\neq 0$, i.e. we can establish
the existence of regular orbits by numerical means even when
the series constructed to represent them are divergent. In
such cases, a low order truncation of the series is often
found to continue to represent some features of the true orbits.

\begin{figure}
\centering
\includegraphics[scale=0.6]{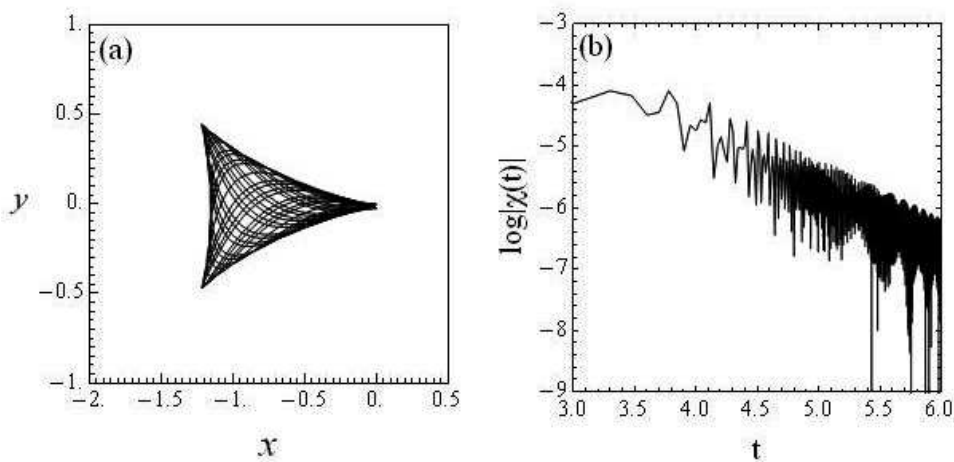}
\caption{(a) A trajectory when $a=b=0.75$, $c=\sqrt{2}/2$,
and $x_0=y_0=0$. (b) The time evolution of the `Finite time
Lyapunov Characteristic Number' $\chi(t)$ (see text) for the
same trajectory, up to the time $t=10^6$. The asymptotic logarithmic
slope is $-1$, indicating that the trajectory is regular.}
\label{hatriangle}
\end{figure}
The regular character of the trajectories can be established
numerically by computing their Lyapunov characteristic numbers.
Such a test is shown in Figure \ref{hatriangle}, referring to a
trajectory in the
model (\ref{eigenharm}) for $a=b=0.75$, $c=\sqrt{2}/2$ and initial
conditions $x_0=y_0=0$. This trajectory was considered in
\cite{coneft2008} (Fig.12c,d of that paper), and it was found
that the series representation of the trajectory clearly diverges.
In order to check whether the trajectory's character remains regular
over longer time intervals, we computed the same trajectory, along
with its variational equations, with a very precise time step
$\Delta t=10^{-5}$ up to a time $t=10^6$. Figure \ref{hatriangle}b
shows the evolution of the `finite time Lyapunov characteristic
number'
\begin{equation}\label{chil}
\chi(t)={1\over t}\ln\mid{\xi(t)\over\xi(0)}\mid
\end{equation}
for this trajectory, where $\xi(t)$ is the length of a deviation
vector computed from the variational equations of motion.
The Lyapunov characteristic number is the limit
$LCN=\lim_{t\rightarrow\infty}\chi(t)$. This limit is
positive for chaotic trajectories, while it is zero for regular
trajectories, for which $\chi(t)$ follows the asymptotic behavior
$\chi(t)\sim t^{-1}$. This is exactly the average power law
found in the case of Fig.\ref{hatriangle}b. Thus, all numerical
indications are that this particular trajectory is regular,
i.e. the domain of initial conditions and parameters leading to
regular trajectories extends beyond the rigorous lower bounds
provided by perturbation theory.

\section{Ordered trajectories far from the center}
Trajectories restricted far from the nodal lines are ordered. For the
trajectories starting in the upper right domain beyond the set of nodal
lines, and close to the diagonal, the initial conditions $x_0,y_0$ are
both absolutely large. Therefore, $1/x_0$, $1/y_0$ are
small. We then set
\begin{equation}\label{xycap}
X(t)={1\over x_0}x(t),~~~Y(t)={1\over y_0}y(t)
\end{equation}
whereby it follows that $X(0)=Y(0)=1$. Assuming that the
trajectories only make small oscillations around the values $X=Y=1$,
we may write expansions of the form
\begin{equation}\label{xyexp}
X(t)=1+X_1(t)+X_2(t)+...,~~~Y(t)=1+Y_1(t)+Y_2(t)+...
\end{equation}
in which we require that the quantities $X_j(t)$, $Y_j(t)$ contain
only terms with coefficients $(x_0^ky_0^l)^{-1}$ satisfying $k+l=j$,
and $X_j(0)=Y_j(0)=0$.

Upon substitution of (\ref{xyexp}) into (\ref{eqmo}), where, in
addition, $x=x_0X$, $y=y_0Y$, the equations of motion for $X,Y$
split into terms of like orders $j$. In the first and second orders
we find the trivial equations
\begin{equation}\label{eqtriv}
{dX_1\over dt}={dY_1\over dt}=0,~~~{dX_2\over dt}={dY_2\over dt}=0
\end{equation}
from which the solutions $X_1=Y_1=X_2=Y_2=0$ are selected. The third
and fourth order  equations read
\begin{equation}\label{eqmo3}
{dX_3\over dt}=0,~~~{dY_3\over dt}={-a\sin ct\over bc^{1/2}y_0^3}
\end{equation}
with solution
\begin{equation}\label{xysol3}
X_3(t)=0,~~~~~~~~~Y_3(t)={a(\cos ct - 1)\over bc^{3/2}y_0^3}~~,
\end{equation}
and
\begin{equation}\label{eqmo4}
{dX_4\over dt}={-\sin(1+c)t\over bc^{1/2}x_0^3y_0},~~~~~~~~
{dY_4\over dt}={a^2\sin 2ct\over b^2cy_0^4}-{\sin (1+c)t\over bc^{1/2}x_0y_0^3}
\end{equation}
with solution
\begin{equation}\label{xysol4}
X_4={\cos(1+c)t-1\over bc^{1/2}(1+c)x_0^3y_0},~~~~~~~~~
Y_4={-a^2(\cos 2ct-1)\over 2b^2c^2y_0^4}+{\cos (1+c)t-1\over bc^{1/2}(1+c)x_0y_0^3}
\end{equation}
respectively. Higher order terms can be found by implementing the recursive
scheme.

\begin{table}
\centering
\begin{tabular}[c]{|c|c|c||c|c||c|c|}
\hline
$x_0=y_0$
& \multicolumn{2}{|c||}{4th approximation}
& \multicolumn{2}{c||}{15th approximation}
& \multicolumn{2}{c|}{numerical} \\
 & ~~$x_{min}^{max}$~~ & $y_{min}^{max}$
 & ~~$x_{min}^{max}$~~ & $y_{min}^{max}$
 & ~~$x_{min}^{max}$~~ & ~~$y_{min}^{max}$~~ \\
\hline\hline
3.4 & 3.4   & 3.4   & 3.412 & 3.4   & 3.412 & 3.4   \\
    & 3.365 & 3.074 & 3.354 & 2.996 & 3.353 & 2.994 \\
\hline
3.2 & 3.2   & 3.2   & 3.215 & 3.2   & 3.216 & 3.2   \\
    & 3.158 & 2.830 & 3.146 & 2.715 & 3.139 & 2.707 \\
\hline
3.0 & 3.0   & 3.0   & 3.021 & 3.0   & 3.022 & 3.0   \\
    & 2.948 & 2.576 & 2.920 & 2.398 & 2.914 & 2.366 \\
\hline
2.8 & 2.8   & 2.8   & 2.830 & 2.8   & 2.869 & 2.8   \\
    & 2.737 & 2.309 & 2.687 & 2.016 & 2.616 & 1.452 \\
\hline
\end{tabular}
\caption{The minimum and maximum values of $x$ and $y$
for some ordered orbits far from the center (initial
conditions $x_0=y_0$ as indicated in the first column),
using a series representation (Eq.\ref{xyexp}) up to
the fourth, or the fifteenth, order, as compared to
numerical results. The last trajectory, for $x_0=y_0=2.8$,
is chaotic. Thus, its series representation is no longer
useful.}
\end{table}
We checked the approximation of the numerical trajectories, when
$a=b=1$ and $c=\sqrt{2}/2$, by the series expansions (\ref{xyexp})
calculated up to order 15 via a computer algebraic program,
We find that for $x_0,y_0$ large enough the approximation is
excellent. However, as we decrease $x_0$ or $y_0$ the deviations between
the analytical and numerical results become appreciable. In fact,
the trajectories of these initial conditions are still ordered and perform
simple oscillations, although the analytical curves by the series
expansions undergo large deviations. The level of approximation,
as we decrease $x_0,y_0$, can be judged from Table I yielding the analytical
maxima and minima of the oscillating functions $x(t)$ and $y(t)$ as derived
from the series truncated at orders $n=4$ and $n=15$. These numbers are compared
with the limits of the trajectories found numerically. In all cases we take $a=b=1$,
$c=\sqrt{2}/2$. For $x_0=y_0\geq 3$ the agreement of both the 4th and 15th
order approximations with the numerical results is very good.

\begin{figure}
\centering
\includegraphics[scale=0.65]{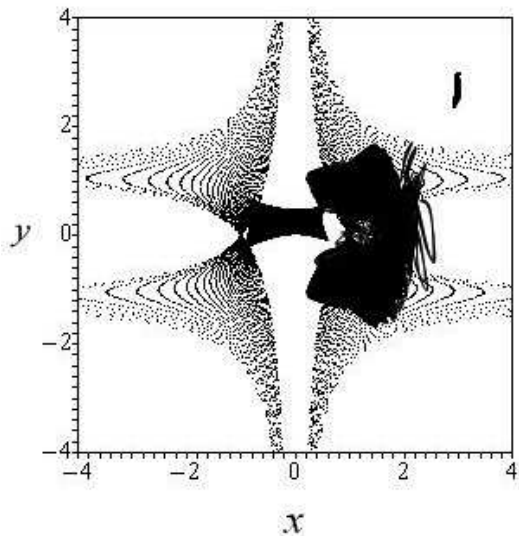}
\caption{An ordered orbit in the inner regular domain (initial
conditions $x_0=y_0=0.5$, a chaotic orbit ($x_0=y_0=1.5$), and
an ordered orbit in the outer regular domain ($x_0=y_0=3.0$)
of the 2D harmonic oscillator model. The trajectories are plotted
on top of the nodal lines.}
\label{hainout}
\end{figure}
On the other hand, for $x_0,y_0$ smaller than $\simeq 2.9$ (but not
smaller than 1) the trajectories come into the domain of the nodal lines
(Fig.\ref{hainout}) and the trajectories become chaotic. This result
shows that the chaotic trajectories are limited in a region covered
by nodal lines (and even then some trajectories are still ordered if
they do not approach the nodal points, as we have seen in the previous
section).

\section{Ordered trajectories in the H\'{e}non - Heiles system}
The main features of the quantum trajectories in the 2D harmonic
oscillator model are maintained to a large extent if we consider
perturbations of this model. Such perturbations present particular
interest in various physical contexts ranging from molecular dynamics
to quantum field theory.

As a simple nonlinear extension of the 2D harmonic oscillator model
we consider the case of the H\'{e}non - Heiles Hamiltonian
\begin{equation}\label{hh}
H=
{1\over 2}(p_x^2+\omega_1^2 x^2)+
{1\over 2}(p_y^2+\omega_2^2 y^2)+
\epsilon x(y^2-{1\over 3}x^2)
\end{equation}
As in \cite{eftcon2006}, we consider the case $\omega_1=1$,
$\omega_2=\sqrt{2}/2$, $\epsilon=0.1118034$.

\begin{figure}
\centering
\includegraphics[scale=0.5]{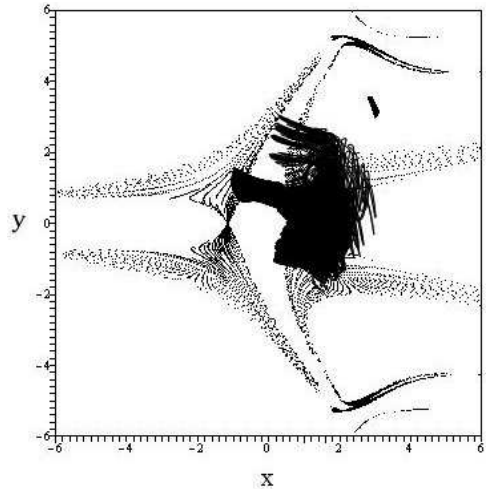}
\caption{Same as in Fig.\ref{hainout}, but for the H\'{e}non-Heiles
model.}
\label{hhinout}
\end{figure}
In order to compute a wavefunction in the above model which, as
in the case of the wavefunction (\ref{eigenharm}) of the 2D harmonic
oscillator model, is the sum of the ground state and two states
with quantum numbers (1,0) and (1,1), we first compute a $200\times 200$
truncation of the Hamiltonian matrix of (\ref{hh}) using the basis
functions of the 2D harmonic oscillator. The matrix
elements considered are given by
\begin{equation}\label{mathh}
H_{ij}=\int_{-\infty}^{\infty}\int_{-\infty}^{\infty}
\phi_i(x,y)\hat{H}\phi_j(x,y)dxdy
\end{equation}
where $\hat{H}$ is the Hamiltonian operator corresponding to
(\ref{hh}) expressed in the position representation, while
$\phi_i(x,y)$ is the i-th eigenstate of the 2D harmonic
oscillator model in the same representation. The index $i$
is related to the 2D harmonic oscillator quantum numbers
$(n,m)$ according to the bijective relation $i=m+1+(n+m)(n+m+1)/2$.
Thus, the i-th eigenvalue of the Hamiltonian (\ref{hh}) satisfies
the relation $E_i=(n+1/2)+(\sqrt{2}/2)(m+1/2)+O(\epsilon)$.
By diagonalizing the truncated real symmetric matrix $\hat{H}_{tr}$,
with entries $H_{ij}$, $i,j=1,200$, we find the eigenvalues $E_i$
and the corresponding normalized eigenvectors $\Phi_i(x,y)$, given by
\begin{equation}\label{eigvechh}
\Phi_{i}=\sum_{j=1}^{200} c^{(i)}_j\phi_j(x,y)
\end{equation}
where $c^{(i)}_j$, $j=1,200$ are the entries of the i-th
column of the orthogonal diagonalizing matrix $C$ satisfying
$C\cdot H_{tr}\cdot C^{-1}=diag_{i=1}^{200}(E_i)$.  Using the
fact that the relation $i(n,m)$ is bijective, for $\epsilon$
sufficiently small we can assign quantum numbers to the i-th
state of the H\'{e}non-Heiles system via the inverse functions
$n(i)$, $m(i)$. Using the above procedure, the eigenvalues and
eigenstates up to $n,m=2$ are computed with an estimated precision
of six significant figures. Furthermore, we denote by $\Phi_{n,m}$
and $E_{n,m}$ the state $\Phi_{i(n,m)}$ as given by Eq.(\ref{eigvechh})
and the corresponding eigenvalue $E_{i(n,m)}$ respectively.

In order, now, to make a comparison of the de Broglie - Bohm
trajectories in the harmonic oscillator and in the H\'{e}non -
Heiles system, we study the trajectories under the wavefunction
\begin{equation}\label{wavehh}
\Psi(x,y,t)=e^{-iE_{0,0}t}\Phi_{0,0}(x,y)+
a e^{-iE_{1,0}t}\Phi_{1,0}(x,y)+
b e^{-iE_{1,1}t}\Phi_{1,1}(x,y)~~.
\end{equation}
We have examined various values of the real parameters $a,b$,
as in the case of the 2D harmonic oscillator model.

Figure \ref{hhinout} shows an example of three Bohmian trajectories
in the above model, i.e. (i) an inner ordered, (ii) a
chaotic, and (iii) an outer ordered trajectory. This figure
compares quite well with the corresponding figure for the
2D harmonic oscillator model (Fig.\ref{hainout}), showing that ordered
Bohmian trajectories persist even for perturbed hamiltonians of
the form (\ref{hh}). In fact, by plotting also the nodal
lines (gray in Fig.\ref{hhinout}), we note that they represent,
to a large extent, a deformation of the nodal lines appearing
in the respective figure (Fig.\ref{haordcha}) for the 2D harmonic
oscillator model. In fact, the only noticeable difference concerns
the right part of the $(x,y)$ square in Fig.(\ref{hhinout}), for
large (positive or negative) values of $y$ and $x>0$. Namely, while
the nodal lines in Fig.\ref{hainout} tend to an asymptote at
$x=0$ for large $y$, the nodal lines in Fig.\ref{hhinout} form
two nearly horizontal zones at $y\approx\pm 4.5$. This, however,
can only affect trajectories very far from the center
(i.e. corresponding to an exponentially small probability),
while the main form of the trajectories, as shown in
Fig.\ref{hhinout}, follows the same features as analyzed in
sections 3 and 4 for the 2D harmonic oscillator model.

\section{Effects on quantum relaxation}

The theory of {\it quantum relaxation}, introduced by Valentini
\cite{val1991}, provides an extension of ordinary quantum mechanics
leading to interesting physical consequences both for quantum particles
\cite{val1991}\cite{valwes2005}\cite{eftcon2006}\cite{ben2010}\cite{towetal2011}
as well as quantum fields \cite{colstru2010}\cite{col2011}. Briefly,
the theory explores the consequences of allowing the possibility
that a set of initial conditions of de Broglie - Bohm trajectories
have an initial probability density $\rho_0$ deviating from the
one postulated by Born's rule i.e. $\rho_0=|\psi_0|^2$, where
$\psi_0$ is the initial wavefunction. The main result, called
`sub-quantum H-theorem' \cite{val1991} predicts that on a coarse-grained
level $\rho$ can only approach closer to $|\psi|^2$ asymptotically in
the forward sense of time (if $\rho_0\neq|\psi_0|^2$), provided that
the initial state has no `fine grained micro-structure', i.e. $\rho$
has initially no great fluctuations on very fine scales (see \cite{val1991}
for more precise definitions). Under such conditions, Valentini defines
an H-function
\begin{equation}\label{hf}
H = \int dq \overline{\rho}\ln(\overline{\rho}/\overline{|\psi|^2})
\end{equation}
where $\overline{\rho}$ is the average particle's density in
cells produced after a choice of coarse-graining in
configuration space, and $\overline{|\psi|^2}$ is the average
value of the square-modulus of the wavefunction in the same cells.
It can be shown that $H$ satisfies the inequality $H(t)-H(0)\leq 0$,
but we emphasize that this does {\it not} imply that $H(t)$ is a
monotonically decreasing function in time. Furthermore, it
is possible to construct states such that $H(t)\simeq 0$ even
if $\overline{\rho}$ differs significantly from $|\psi|^2$ from
cell to cell, since the integrand in (\ref{hf}) can be both
positive or negative. However, simulations indicate that the
quantum relaxation is a numerical fact in several cases. In
subsequent calculations, we probe directly the approach of
$\rho$ to $|\psi|^2$ on the basis of an alternative indicator,
namely the integral of the absolute difference $|\rho-|\psi|^2|$
over the whole configuration space (see below), discussing also
the relevance of this test to tests based on the temporal behavior
of the H-function.

The interest in the quantum-relaxation theory is twofold:
i) it provides a dynamical basis for deriving (rather than postulating)
Born's rule, and ii) it leads to interesting new physics if we consider
that fluctuations from Born's rule have a considerable amplitude during
the early stages of evolution of particular quantum systems. A specific
implementation regards the dynamical behavior of quantum fluctuations
in the early Universe \cite{petpin2008}\cite{val2010}. In fact,
it was found  \cite{val2010} that if there were deviations from
{\it quantum equilibrium} (i.e. Born's rule) in the early Universe,
these could have led to detectable imprints in some observables
encountered in the framework of physical cosmology (e.g. in the CBR
spectrum).

Concrete numerical simulations of the relaxation mechanism
were produced in \cite{valwes2005}, \cite{eftcon2006}, and
\cite{towetal2011}.
In all cases, it was found that the rate at which the
phenomenon occurs depends on the existence of {\it chaotic}
quantum trajectories, due to the appearance of {\it quantum
vortices}, i.e. one or more vortex structures of the quantum
flow. On the other hand, in \cite{eftcon2006}
we noted that the presence of {\it ordered} trajectories
suppresses quantum relaxation. Here, we explore this
phenomenon in greater detail, but we also give an example
of a rather surprising phenomenon, namely that quantum
relaxation can be suppressed even in some cases of ensembles
of chaotic rather than regular trajectories.

\begin{figure}
\centering
\includegraphics[scale=0.65]{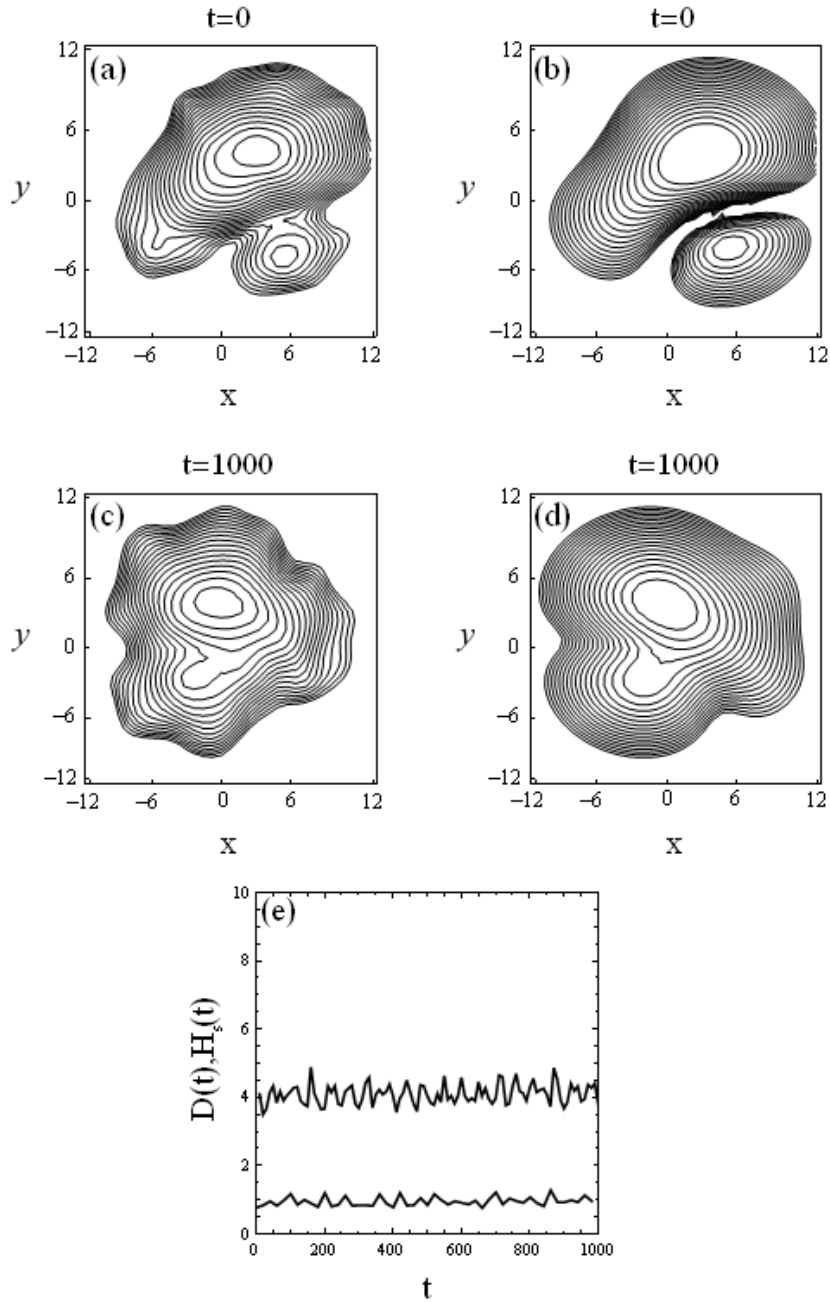}
\caption{(a) The contour plots of the density $P_s(x,y,t)$ in
a test case, where 961 particles are given initial conditions
so as to represent a discrete realization of the probability
$|\psi|^2$ given by Eq.(\ref{eigenharm}) for $t=0$. (b) Contour
plots of the exact density $\rho=|\psi|^2$ at $t=0$.
(c) and (d) Same as in (a) and (b), but for $t=1000$.
The particles positions are now given by evolving their quantum
trajectories. (e) Time evolution of the density difference $D(t)$
(upper curve) and of the H-function $H_s(t)$ (lower curve) in
the same simulation. }
\label{reltest}
\end{figure}
\begin{figure}
\centering
\includegraphics[scale=0.6]{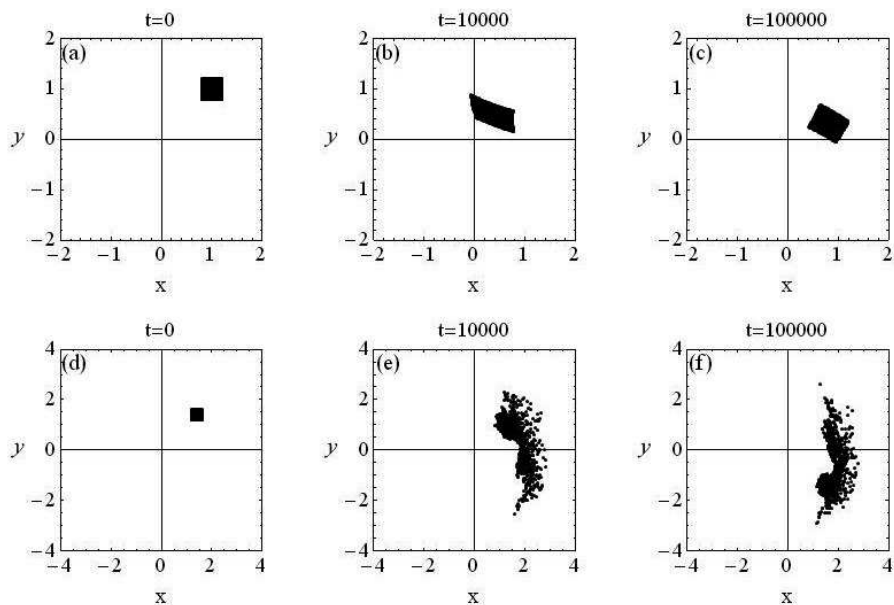}
\caption{(a)The initial conditions of 961 particles in the
`ordered' ensemble of quantum trajectories considered for relaxation
effects(see text) in the model of Eq.(\ref{eigenharm}). (b),(c)
Positions of the same particles at the times $t=10000$ and
$t=100000$ respectively. (d),(e),(f) Same as in (a),(b),(c),
but for a `chaotic' ensemble of quantum trajectories.
}
\label{traordcha}
\end{figure}
\begin{figure}
\centering
\includegraphics[scale=0.5]{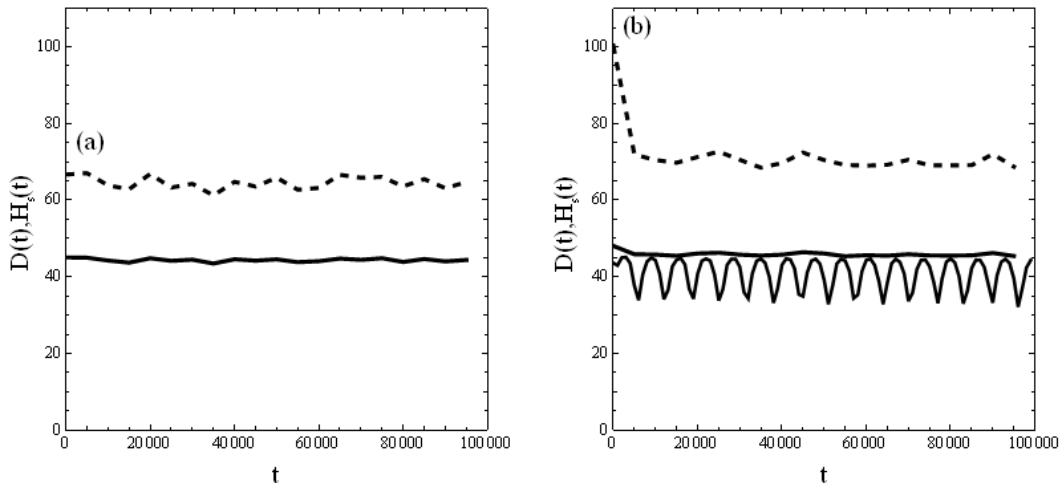}
\caption{The time evolution of the density difference $D(t)$
(solid curve) and of the coarse-grained H-function $H_s(t)$
(dashed curve) for the ensemble of (a) the upper panels and
(b) the lower panels of Fig.\ref{traordcha}. The lower
oscillatory curve in (b) refers to a calculation of $D(t)$
in only the right semi-plane $x>0$, denoted by $\overline{D}(t)$.
}
\label{relordcha}
\end{figure}

Figures \ref{reltest},\ref{traordcha},\ref{relordcha} show the
basic result. Figure 7 shows first a `test' case, in which we
compute the Bohmian trajectories of 961 (=31$\times$31) trajectories
in the 2D harmonic oscillator model (\ref{eigenharm}), with $a=b=1$,
$c=\sqrt{2}/2$. The initial conditions are chosen by a rejection
algorithm (see \cite{pressetal1986}) so as to provide a discrete
sampling of an initial density distribution $\rho_0=|\psi_0|^2$.
We consider a relatively small number of particles (of order 
1000) because we require that their trajectories should be
subsequently computed by a small (and hence time-consuming) timestep
$\Delta t=10^{-4}$. This is needed in order to guarantee that no
spurious numerical effects are introduced at least for the 
{\it regular} trajectories, which are our main concern here. 
In fact, for some trajectories we checked the results with 
a fixed timestep against an integration with variable time 
step supplemented by a regularization scheme when the distance
from a nodal point becomes smaller than unity, as developed in
\cite{eftetal2009}, and found that both integrations are consistent.

The above limitation regarding the number of particles notwithstanding,
we can compute a smooth approximation of the coarse-grained density
sampled by these trajectories via the formula used in \cite{eftcon2006}:
\begin{equation}\label{prosm}
P_s(x, y, t)=\sum_{i=1}^{961}
A\exp\left[{(x-x_i(t))^2 + (y-y_i(t))^2\over 2\sigma^2}\right]
\end{equation}
where $\sigma$ is a Gaussian smoothing length, set equal to
$\sigma=0.3$. The constant $A$ is computed so that the
integral of the function $P_s(x,y,t)$ over the whole
configuration space is equal to unity. Figures 7a,b display
a comparison of the contour plots of the density $P_s(x,y,0)$
(Fig.7a) and of $|\psi(x,y,0)|^2$ (Fig.7b), showing that,
by this choice of initial conditions, $P_s$ turns
to approximate fairly well the density $|\psi(x,y,0)|^2$.
The degree of approximation can be judged by computing (in
time) an overall absolute density difference measure $D(t)$
in a grid of points, given by
\begin{equation}\label{dendif}
D(t)=\sum_{k=1}^N\sum_{l=1}^N
\left| Ps(x_k,y_l,t) - |\psi(x_k,y_l,t)|^2\right|
\end{equation}
where, as in \cite{eftcon2006}, we set
$N=128$ and $x_k=-N/10+k/5$,$y_l=N/10+l/5$ in the sum
(\ref{dendif}). The initial value of this quantity for the
above sample of orbits turns to be $D(0)\simeq 4$. As shown
in \cite{eftcon2006}, such a value is compatible with Poisson
noise deviations of $P_s(0)$ from $|\psi_0|^2$, i.e. the two
distributions are equal to within statistical fluctuations.
Now, since the time evolution of the density resulting from
a set of Bohmian trajectories should respect the continuity
equation, as we evolve the Bohmian trajectories we should
find this equality being preserved at all times $t$. This is
indeed observed by a comparison of the contour plots of $P_s$
and $|\psi|^2$ at subsequent times, as in Figs.7c,d, where
this comparison is shown for the time $t=1000$. The time
evolution of $D(t)$ up to $t=1000$ is displayed in Fig.7e,
showing that $D(t)$ remains stuck at the statistical noise
level $D(t)\simeq 4$, i.e. no significant deviations
of $P_s$ from $|\psi|^2$ develop as the numerical
integration of the Bohmian trajectories proceeds in
time. In the same plot, we show the behavior of the
function
\begin{equation}\label{hsig}
H_s(t)=\sum_{k=1}^N\sum_{l=1}^N
P_s(x_k,y_l,t)\log\left(P_s(x_k,y_l,t)/|\psi(x_k,y_l,t)|^2\right)
\end{equation}
which is similar to Valentini's coarse-grained H-function
computed over a coarse-graining at a resolution scale
$\sim\sigma$. In fact, in the present as well as in all subsequent
examples, we find that the two functions $D(t)$ and $H_s(t)$
have a quite similar temporal behavior. This is because our
way to choose initial conditions localized in small boxes
implies that the main contribution to the sum (\ref{hsig})
comes from cells (labeled by $(k,l)$ where one has
$P_s>|\psi|^2$, whence it follows that
$$
P_s\log(P_s/|\psi|^2) = P_s-|\psi|^2 + O[P_s(P_s/|\psi^2| -1)^2]~~.
$$
We stress that, while this alternative calculation aims to
approximate numerically the temporal behavior of the coarse-grained
H-function, the most direct numerical test of the closeness of $\rho$
to $|\psi|^2$ is provided by the numerical value of the quantity
$D(t)$.

Figure \ref{traordcha}, now, shows what happens if, instead, we start
with initial conditions such that $P_s\neq |\psi|^2$. Figures
\ref{traordcha}a,b,c show the instantaneous positions at the times
$t=0$ (Fig.\ref{traordcha}a), $t=10000$ (Fig.\ref{traordcha}b) and
$t=100000$ (Fig.\ref{traordcha}c), of 961 trajectories
with initial conditions taken in a uniform $31\times 31$ grid
in the square box $(0.9,1.1)\times(0.9,1.1)$, in the same
2D harmonic oscillator model as in Fig.\ref{reltest}. The central
trajectory with initial conditions $x_0=y_0=1$ is regular
(Figure \ref{haordcha}a). The same holds true for all its surrounding
trajectories in the square. In Figs.\ref{traordcha}a,b,c it is
then observed that the trajectories travel all together without
any phase mixing taking place. As a result, the square area
formed by this grid moves coherently from one to another
domain of the configuration space, without exhibiting any
systematic stretching or contracting along particular directions.
This behavior is due to the fact that the numerical trajectories
are found to exhibit quasi-periodic oscillations with two
incommensurable frequencies which are the same for all the
trajectories, namely $\omega_1=1$, and $\omega_2=c$. In fact,
this behavior is also suggested by the series solutions of
perturbation theory found for smaller values of $a$, and $b$,
which share the same property. As a result, no phase mixing
is possible for such trajectories.

Figures \ref{traordcha}d,e,f show now a similar calculation for
961 trajectories with initial conditions in the square box
$(1.3,1.5)\times(1.3,1.5)$. In this case, the central trajectory
($x_0=y_0=1.4$) as well as all surrounding trajectories are
chaotic. We now see that the trajectories expand in a
considerable domain on both sides of the x-axis, however they
always reside in the {\it same semi-plane} (i.e. right) with
respect to the y-axis, i.e. the orbits never cross the y-axis.
The no-crossing of the y-axis is due
to the central gap formed between the innermost hyperbola-like
limit of the nodal lines on both sides of the y-axis (cf. Fig.
\ref{haordcha}) The presence of such limits was shown
theoretically in EKC. In fact, we have integrated some
representative trajectories in the same sample up to a time
$t=10^5$, using a precise time step ($\Delta t=10^{-4}$),
and no hint of crossing of the y-axis was observed for any
of them.

The fact that there can be chaotic trajectories limited within only
a sub-domain of the total available configuration space implies
that obstructions to quantum relaxation can be present even under
chaotic de Broglie - Bohm dynamics. This is shown in Fig.\ref{relordcha},
showing a comparison of the evolution of $D(t)$ (Eq.(\ref{dendif}))
for (a) the set of regular trajectories (upper panels in Fig.\ref{traordcha}),
and (b) the set of chaotic trajectories (lower panels in Fig.\ref{traordcha}).
In the case of regular trajectories (Fig.\ref{relordcha}a), we find that
$D(t)$ undergoes small oscillations around a mean value $D(t)=45$. This
is consistent with the fact that ordered trajectories suppress quantum
relaxation. However, the {\it same} trend is observed in the case of the
chaotic trajectories, as shown in Fig.\ref{relordcha}b, since these
trajectories never fill the left semi-plane $x<0$.

In this case also the value of $D(t)$ undergoes small oscillations
around $D(t)\simeq 45$, and it does not tend to $D(t)=0$ as $t$
increases. But it should be pointed out that important deviations
of $P_s$ from $|\psi|^2$ are observed even if we limit ourselves
to considering the sum (\ref{dendif}) taken only on the semi-plane
$x>0$ (in this case we normalize both $P_s$ and $|\psi|^2$ so that
the surface integral of both quantities over the semi-plane $x>0$
is equal to unity). We observe now that the new curve (denoted by
$\overline{D}(t)$) undergoes oscillations, starting from a value
$\overline{D}=45$ and falling at particular moments down to
$\overline{D}=35$. Such trend is, however, subsequently reversed,
and there are many cycles of rise and fall of $\overline{D}(t)$,
but $\overline{D}(t)$ does not tend to zero.

A qualitative explanation of this behavior is provided by noting that
the form of the wavefunction (\ref{eigenharm}) allows for a recurrent
transfer of quantum probability mass from the right to the left
semi-plane and vice-versa. However, the considered ensemble, of only
chaotic trajectories, contains no trajectories accounting for such
transfer, i.e. no trajectories oscillating from one semi-plane to
the other, but this transfer is only achieved in  ensembles
containing regular trajectories. Thus, despite the fact that we have
chaotic trajectories, in this example $P_s$ cannot tend to $|\psi|^2$
as $t$ increases. We emphasize that this phenomenon occurs despite
the fact that the wavefunction in the present case contains no
disjoint supports. On the other hand, if there are two or more
disjoint supports of the wavefunction, if we take initial conditions
in only one support, relaxation never occurs even if all the
trajectories are chaotic, since such trajectories cannot pass
from the domain of one support to that of another at any later
time $t$.

\section{Further examples of obstruction to relaxation.
Quantitative estimates}

\subsection{Further examples}
\begin{figure}
\centering
\includegraphics[scale=0.7]{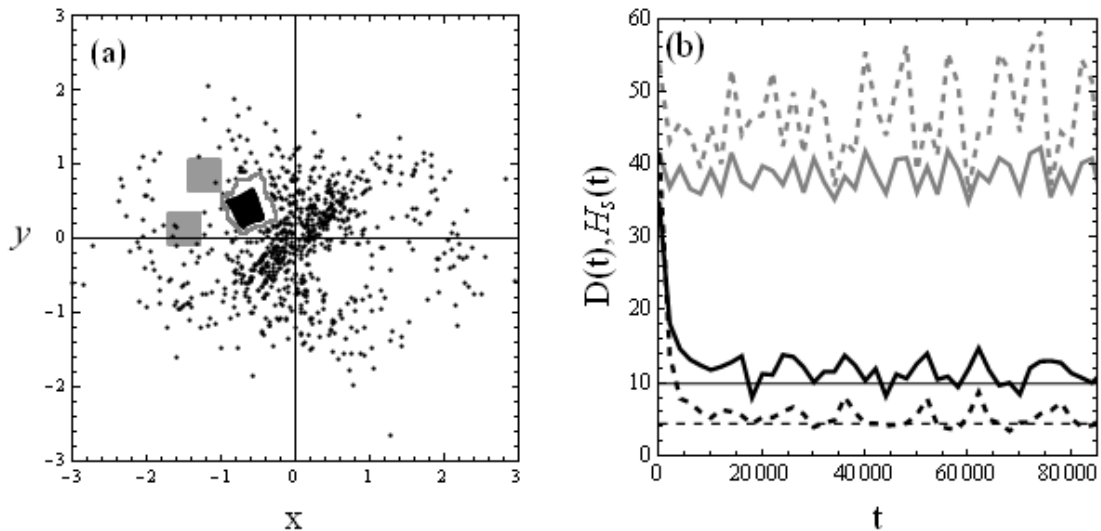}
\caption{(a) The gray square boxes show the domains of initial conditions
for a set of 961 trajectories in the wavefunction model (\ref{psi002011}),
around the values (i) i) $x_0=-1.5$, $y_0=0.1275$ (bottom square) and ii)
$x_0=-1.23$, $y_0=0.84$ (up square). The black box contains the images
of the points of the first box, at $t=10000$, while the scattered points
are the images of the points of the second box at the same time.
The closed gray curve shows the approximate limit of the regular
domain at the same time. (b) Time evolution of the quantities $D(t)$
(solid), and $H_s(t)$ (dashed) for the ensemble (i) (top curves, gray,
regular trajectories), and (ii) (lower curves, black, chaotic trajectories).
The straight solid line shows an average `noise' level for $D(t)$ when we
consider initial conditions satisfying $\rho_0=|\psi_0|^2$.
The curve $D(t)$ for the ensemble (ii) is, on average, a little above
this level  by a value $\Delta D\simeq 2.4$, which is consistent with
a calculation  taking into account the area left empty by the chaotic
trajectories. }
\label{rel002011}
\end{figure}

The examples treated in sections 2 and 5 are rather special
because even the chaotic trajectories are restricted
on parts only of the total available configuration space. But
we can construct other examples of mixed (i.e. co-existing
regular and chaotic) dynamics where such restriction no longer
holds. One such case is provided by the wavefunction model
\cite{eftetal2009}:
\begin{equation}\label{psi002011}
\psi(x,y,t))=e^{-{x^2+cy^2\over 2}}e^{-{1+c\over 2}it}
\left[1+a(x^2-1)e^{-2it}+bc^{1/2}xye^{-it}\right]
\end{equation}
for $a,b,c$ real. Below, we choose $a=1.23$, $b=1.15$, $c=\sqrt{2}/2$.
In \cite{eftetal2009}it was shown that this model contains both ordered
and chaotic orbits. Figure \ref{rel002011} shows a calculation of two
distinct sets of 961 trajectories each taken with initial conditions
inside a $0.4\times 0.4$ box centered at i) $x_0=-1.5$, $y_0=0.1275$
and ii) $x_0=-1.23$, $y_0=0.84$. The central values (i) and (ii)
correspond to a regular and a chaotic orbit respectively. The main
remark concerns the fact that, as is the case of regular orbits of
the model of sections 2,3, and 5, the regular orbits in the present
model undergo no phase mixing as well, i.e. they all do quasi-periodic
oscillations with the frequencies $\omega_1=1$
and $\omega_2=c=\sqrt{2}/2$. Furthermore, the boundary of the
area occupied by the regular trajectories cannot be penetrated
by chaotic orbits. This effect is shown in Fig.\ref{rel002011}a
depicting the initial conditions of the ordered and chaotic sets
at $t=0$, and their images at $t=10000$, i.e. a time after which
the chaotic trajectories spread over most of their allowable
space belonging to the support of the system's wavefunction,
while the ordered trajectories remain inside the black box.

Figure \ref{rel002011}b
shows the relaxation plots for the ensembles (i) and (ii)
respectively. In the case of the ensemble (i) (regular orbits),
we notice the clear deviation from relaxation, while in the case
of the ensemble (ii) (chaotic orbits), both curves $D(t)$ and
$H(t)$ approach a limiting mean value $D\simeq 12.3$, and
$H_s\simeq 6.4$. In fact, we notice (Fig.\ref{rel002011}a)
that the chaotic trajectories do not penetrate the boundary
(closed gray curve) corresponding to the set of regular
trajectories, thus, while $|\psi|^2$ covers all available space,
the value of $\rho(t)$ is close to zero at all times
within the domain delimited by this boundary. By calculating
the percentage of the area occupied by regular orbits, versus
the total area where all the orbits spread, we then estimate
that the values reached by $D(t)$ and $H(t)$ on average show 
the expected difference from the 'noise' threshold indicating
complete relaxation, which, in this case, is found to be
$D=9.9$ and $H_s=4.4$ respectively.

\begin{figure}
\centering
\includegraphics[scale=0.7]{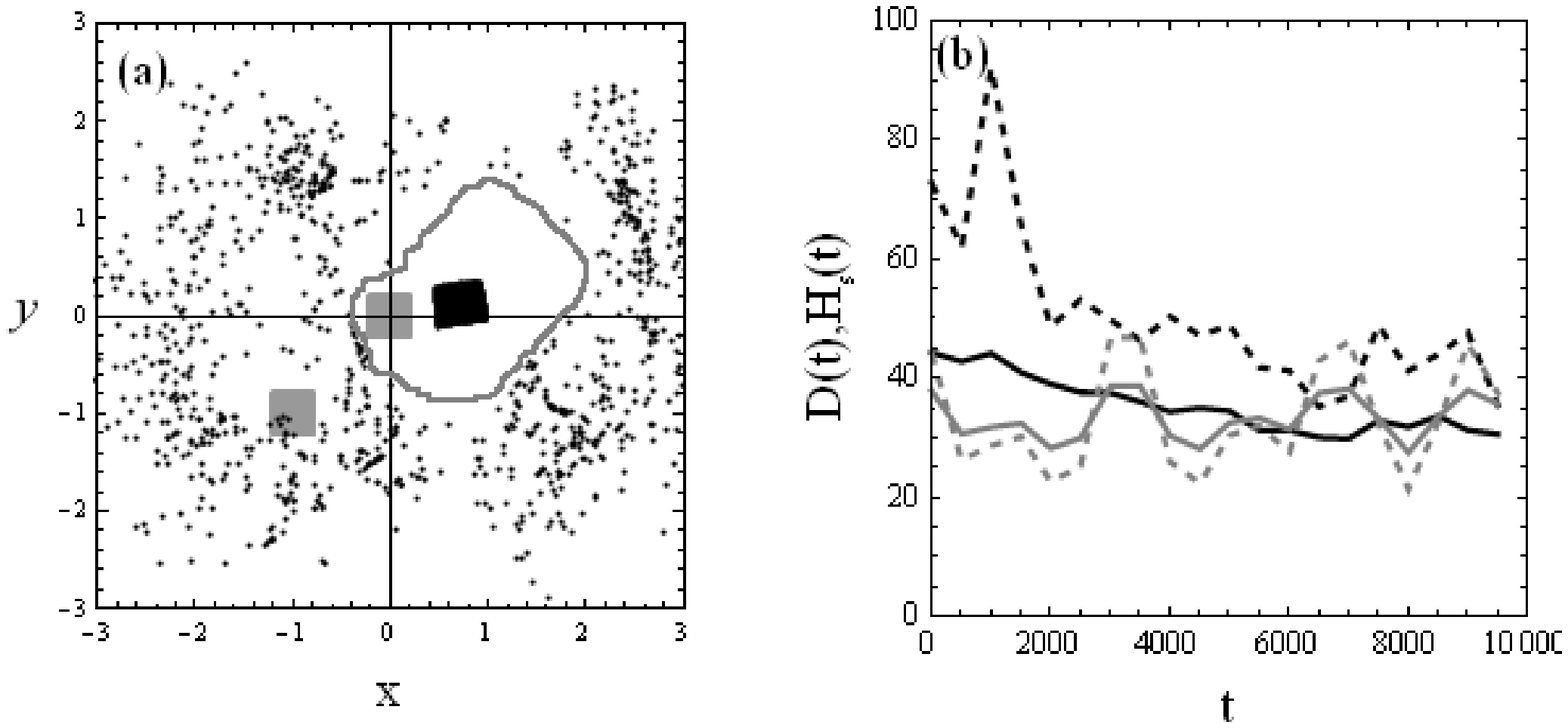}
\caption{Same as in Fig.\ref{rel002011}, but for a superposition
of four eigenfunctions (Eq.(\ref{psi00102011}). In this case, the
initial square boxes are centered around (i) $x_0=y_0=0$ (regular
trajectories), and (ii) $x_0=y_0=-1.1$ (chaotic trajectories).
In (b) we note the large values of $D(t)$ and $H_s(t)$ for the
chaotic ensemble as well, due to the fact that the chaotic trajectories
are excluded from a large domain of the configuration space.}
\label{rel00102011}
\end{figure}
Similar phenomena are found if we consider a superposition of four
eigenfunctions
\begin{eqnarray}\label{psi00102011}
\psi(x,y,t))&=&e^{-{x^2+cy^2\over 2}}e^{-{1+c\over 2}it}\nonumber\\
&\times&
\left[1+a(x^2-1)e^{-2it}+bc^{1/2}xye^{-i(1+c)t}+dxe^{-it}\right]
\end{eqnarray}
with $a=b=d=1$, and $c=\sqrt{2}/2$ (Figure \ref{rel00102011}).
In fact, in this case, the domain occupied by regular orbits is
numerically found to be larger than in the case of the wavefunction
model (\ref{psi002011}) which is the superposition of a smaller
number of states. In general, there appears to be no simple rule
yielding the extent of the domain of regular orbits as the number
of superposed eigenfunctions increases.

In systems like the 2D harmonic oscillator (Eq.(\ref{ham2dharm})
with $c$ rational), or in the case of the square box examined e.g.
in \cite{valwes2005},
the following property can be readily shown to hold true: an arbitrary
combination of any number of eigenfunctions with real amplitudes
implies that all motions are periodic (see \cite{conetal2008}).
In such systems, $D(t)$ and the coarse grained H-function are
both periodic functions of time i.e. we have $D(T)=D(0)$ and
$H(T)=H(0)$, where $T$ is the period. One dimensional periodic
models exhibiting the same effect were considered in \cite{val2001}.
In fact, this criterion can be used as a test of the accuracy
of numerical integrations, i.e. by checking the periodicity of the
numerically computed functions $D(t)$ or $H(t)$ in a periodic model.

If, now, the frequency ratio $c$ in the Hamiltonian (\ref{ham2dharm})
is irrational, $H(t)$ has no periodic behavior. However, there
can still be defined approximate periods by considering, for
example, the rational truncates $q_n/p_n$ of the continued
fraction representation of the frequency ratio $c=\omega_2/\omega_1$
(assuming, without loss of generality, $\omega_1>\omega_2$, i.e.
\begin{equation}\label{cffreq}
\omega_2/\omega_1 = [a_1,a_2,a_3,...]=
{1\over {a_1+{1\over a_2+{1\over a_3+\ldots}}}}
\end{equation}
The n-th rational truncate, corresponding to the rational number
$q_n/p_n=[a_1,\ldots,a_n]$, defines an approximate period
$T_n\approx 2\pi p_n/\omega_1$. Furthermore, for a large measure
of incommensurable frequency pairs the following diophantine
relation holds:
$$
|q_n\omega_1-p_n\omega_2|\approx {\gamma\over q_n+p_n}
$$
for a positive constant $\gamma$. On the other hand, the de Broglie-Bohm
equations of motion can be written as combinations of odd trigonometric
functions with arguments $(n_1\omega_1+n_2\omega_2)t$. Then, using
$\omega_1=p_n\omega_2/q_n+O[1/(q_n(q_n+p_n))]$ and expanding the
equations of motion we find that at the times $T_n\approx 2\pi p_n/\omega_1$
{\it all} trajectories are recurrent to nearly their initial positions
to within an accuracy $O[2\pi/(p_n+q_n)]$. Since both $q_n,p_n$ increase
with $n$, we conclude that all trajectories come arbitrarily close to
their initial conditions at very specific (and quite long, as $n$
increases) times. This phenomenon can be considered as a quantum
analog of the Poincar\'{e} recurrence for the de Broglie - Bohm
trajectories.

\subsection{Hamiltonian case ($a,b$ complex)}

\begin{figure}
\centering
\includegraphics[scale=0.5]{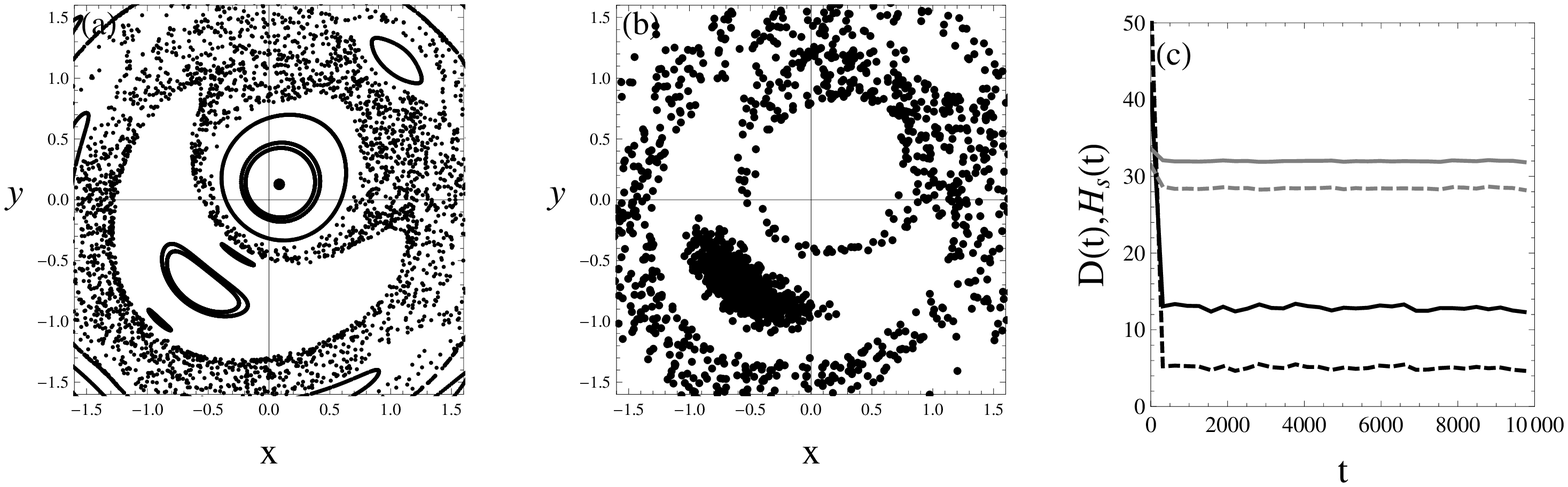}
\caption{(a) Stroboscopic surface of section of the de Broglie - Bohm
trajectories in the model (\ref{wispuj}) with $a=0.17651$, $b=d=1$,
$\gamma_1=3.876968$, $\gamma_2=2.684916$, produced by taking the iterates 
up to $t=5000$ of each one of 15 initial conditions along the diagonal 
$x=y$ from $x=y=-1.6$ to $x=y=1.6$. (b) The iterates, at time
$t=2000\pi$, of a set of 961 initial conditions in a square box
$0.2\times 0.2$ centered at $x_0=y_0=-0.6$ (regular) remain constrained
in the lower left island of stability, while the iterates of
the initial conditions in a similar box around $x_0=y_0=0.9$
(chaotic) fill a chaotic domain. (c) Time evolution of $D(t)$ (solid
curves) and $H_s(t)$ (dashed curves) for the regular (top curves)
and chaotic (lower curves) trajectories of (b). }
\label{wispujrel}
\end{figure}
The equality of the frequencies for all regular orbits is a particular
property of the de Broglie - Bohm dynamics in systems with a superposition
of three or more eigenfunctions with a real amplitude. On the other
hand, when the superposition amplitudes are complex rather than real,
a novel feature may appear, namely we can construct particular
combinations leading to a Hamiltonian representation of the de Broglie
- Bohm equations of motion. In such cases the regular trajectories
still exhibit quasi-periodic oscillations, but the frequencies are
not fixed, i.e. they are not the same for all trajectories.  However,
obstructions to relaxation can still be posed by the appearance of
{\it islands of stability} in the configuration space.

An example of this phenomenon is provided by the model considered in
\cite{wispuj2005}:
\begin{equation}\label{wispuj}
\psi(x,y,t)=e^{-{x^2+y^2\over 2}}\left(a e^{-it}+bxe^{-2it-\gamma_1}
+dye^{-2it-\gamma_2}\right)
\end{equation}
with $a=0.17651$, $b=d=1$, $\gamma_1=3.876968$, $\gamma_2=2.684916$.
The wavefunction (\ref{wispuj}) corresponds to a superposition of the
states $\psi_{00}$,$\psi_{10}$ and $\psi_{01}$ in the Hamiltonian
(\ref{ham2dharm}) with $c=1$. In this particular case the de
Broglie - Bohm equations of motion are given by a time-dependent
Hamiltonian model \cite{wispuj2005}, with period $T=2\pi$, where
$y$ is equivalent to a momentum variable canonically conjugated to
the position $x$. If we plot $x(t),y(t)$ along particular trajectories
at the period multiples $t=2\pi,4\pi,...$, we obtain stroboscopic
surfaces of section presenting typical features of a Hamiltonian
system with mixed phase space. In Fig.\ref{wispujrel}a we see this
structure, which consists of the co-existence of several islands of
stability surrounded by an intricate structure and chaotic layers.
The key remark regarding relaxation is that the regular orbits
cannot escape from the boundaries of the islands of stability.
Thus, if we start with initial conditions exclusively in the domain
occupied by one or more islands, the particles' coarse-grained
density $\overline{\rho}$ always retains the value $\overline{\rho}(t)=0$
in the chaotic domain of the configuration space, and vice versa, i.e.
initial conditions in the chaotic domain lead to $\overline{\rho}(t)=0$
inside the islands. This effect is shown in Fig.\ref{wispujrel}b.
The consequences for relaxation are shown in Fig.\ref{wispujrel}c,
where we observe that both quantities $D(t)$ and $H_s(t)$ remain far
from zero for all times $t$ up to $t=10000$. We note that the
regular trajectories inside an island (like the large lower left 
island in Fig.\ref{wispujrel}a) exhibit a local phase mixing, which 
leads to a uniform distribution with respect to the angles after 
sufficiently long time.  In fact, inside an island
we can define a local form of an approximate integral of motion
$I(x,y,t)$ (see \cite{con1960}), i.e. a formal series in $x,y$
with time-dependent trigonometric coefficients, whose formal time
derivative, depending on the Poisson bracket with the Hamiltonian
$H$, is given by an exponentially small estimate:
\begin{equation}\label{intepois}
{dI\over dt}={\partial I\over\partial t}+\{I,H\}=O(\exp(-1/r))
\end{equation}
where $r=[(x-x_0(t))^2+(y-y_0(t))^2]^{1/2}$, and $(x_0(t),y_0(t))$
is the time evolution along the periodic orbit at the center of the
island of stability (see \cite{eftetal2004} for a heuristic
derivation of such estimates). This implies that, down to an
exponentially small error, we can practically consider the
invariant curves within an island of stability of Fig.\ref{wispujrel}a
as given by the level curves of $I(x,y,0)=C$, for various values of
the constant $C$. Then, it is straightforward to show that if we
partition the configuration space within an island by considering
a sequence of level values of the parameter $C$, then, independently
of how we partition the remaining configuration space, the coarse-grained
H-function remains bound away from zero at all times $t$.

Finally, we should mention the cases of {\it coherent state}
wavefunctions (see our work \cite{eftcon2006}), consisting of
a superposition of infinitely many eigenfunctions, which exhibit
no relaxation and contain only regular trajectories.

In view of the above examples, we conclude that order in de
Broglie - Bohm quantum mechanics appears in a quite significant
variety of contexts. On the other hand, in some other cases
(e.g.\cite{valwes2005}\cite{towetal2011}), using a larger number
of eigenfunctions, relaxation effects were found to be complete.
But in the case of exact or approximate coherent states,
there can be all-time or finite-time obstructions to relaxation
which, as found in \cite{eftcon2006} persist for times exponentially
long in the inverse of a parameter (equivalent to the system's
mean energy). The fact that such states are superpositions of
infinitely many eigenfunctions shows that the number of
eigenfunctions composing a particular quantum state cannot be
used always as a good indicator of the speed of quantum relaxation.
The issue of which particular superposition states speed up or
slow down the relaxation process is interesting and is left for
future research.

\section{Conclusions}

We examined the role of order in the de Broglie - Bohm approach
to quantum mechanics, by studying the conditions for the
existence of regular quantum trajectories, i.e. trajectories
with a zero Lyapunov characteristic number. We found also a
partial order even in some cases of chaotic trajectories.
Finally, we studied the consequences of order in the phenomenon
of quantum relaxation. Our main conclusions are the following:

1) We demonstrated, via low-order perturbation theory, that there
are trajectories avoiding close encounters with the moving nodal
points, in an example of a quantum state consisting of a superposition
of three eigenfunctions in the 2D harmonic oscillator model.
Due to this effect, there can be regular quantum trajectories
extending spatially in a domain overlapping with the domain
of nodal lines.

2) We computed series expansions for regular trajectories both
in the interior and the exterior of the domain covered by
nodal lines. We provided an estimate of the domains of the
series convergence, and compared this with numerical trajectories.
It was found that the series only provide a lower bound of the
domains where regular trajectories are calculated by numerical
means.

3) We examined the quantum trajectories in the H\'{e}non - Heiles
system, for quantum states similar to the ones used in the
2D harmonic oscillator models. The two cases yield a qualitative
agreement as regards the extent of the domain and the initial
conditions leading to regular trajectories. In the case of
the H\'{e}non - Heiles model, however, the nodal lines develop
new structures leading to more chaotic trajectories.

4) We studied the influence of order in the so-called {\it
quantum relaxation} effect, i.e. the approach in time of a
spatial distribution $\rho$ of particles following quantum
trajectories to Born's rule $\rho=|\psi|^2$, even if initially
$\rho_0\neq|\psi_0|^2$. In previous works, it was identified
that chaos provides in general the dynamical substrate leading
to the realization of Born's rule. Here, however, we show
cases where the relaxation can be effectively suppressed
when we consider ensembles not only of regular, but also of
chaotic quantum trajectories. This is due to the fact that
the existence of order in a system poses obstructions
in the extent to which chaotic trajectories can mix in the
configuration space. For the quantification of relaxation
effects we compute two different quantities, namely a
density difference $D(t)$ and an H-function $H_s(t)$ over
a smoothed particle density $P_s$, and show that both quantities
exhibit a similar temporal behavior.

5) We examined several models of superposition of a small
number (three or four) of eigenfunctions and found that 
the existence of order in such models is generic. We also
examined the case of complex superposition amplitudes,
and explained how the appearance of islands of stability
creates a new mechanism of obstruction of quantum relaxation.
The degree of generality of the above phenomena is proposed
as a subject for future study.\\
\\
\noindent
{\bf Acknowledgments:} C. Delis was supported by the State Scholarship
Foundation of Greece (IKY) and by the Hellenic Center of Metals
Research. C.E. has worked in the framework of the COST Action
MP1006 - Fundamental Problems of Quantum Physics.

\section*{References}

\end{document}